\documentclass[12pt]{article}
\usepackage[dvips]{graphics}
\usepackage{graphicx}
\usepackage{dcolumn}
\usepackage{bm}
\usepackage{amsmath}
\usepackage{amssymb}
\usepackage{color}
\definecolor{darkblue}{rgb}{0,0,.5}
\definecolor{darkgreen}{rgb}{0,0.5,0}
\definecolor{darkred}{rgb}{0.5,0,0}
\usepackage[colorlinks = true,
            linkcolor = darkblue,
            urlcolor  = darkblue,
            citecolor = darkblue,
            anchorcolor = blue]{hyperref}

\newcommand{\beq}{\begin{equation}}
\newcommand{\eeq}{\end{equation}}
\newcommand{\beqa}{\begin{eqnarray}}
\newcommand{\eeqa}{\end{eqnarray}}

\begin{document}
\def\ii{\'\i}

\title{The black hole merger event GW150914 within a modified theory of
General Relativity}

\author{Peter O. Hess \\
{\small\it
Instituto de Ciencias Nucleares, UNAM, Circuito Exterior,}\\
{\small\it
 C.U., A.P. 70-543, 04510, Mexico D.F., Mexico} \\
{\small\it and} \\
{\small\it 
Frankfurt Institute for Advanced Studies, Wolfgang Goethe University,}\\
{\small\it 
Ruth-Moufang-Strasse 1, 60438 Frankfurt am Main, Germany}
}


\maketitle

\begin{abstract} 
In February 2016 the first observation of gravitational waves
were reported. The source of this event, denoted as GW150914, was identified as
the merger of two black holes with a about 30 solar masses each, at a distance
of approximately 400Mpc. These data where deduced using the Theory of General Relativity.
Since 2009 a modified theory was proposed which adds near massive objects 
phenomenologically the contribution of a dark energy, whose origin are vacuum fluctuations.
The dark energy accumulates toward smaller distances, reducing effectively
the gravitational constant. In this contribution we show that as a consequence the
deduces chirping mass and the luminosity distance are larger. This
result suggests that the black hole merger corresponds to two massive black holes 
near the center of primordial galaxies at large luminosity distance, i.e. large
redshifts.
\end{abstract}

\section{Introduction}
\label{intro}

In February 2016 the first observation of gravitational waves was reported by the
LIGO collaboration \cite{Ligo1} and  in \cite{Ligo2} more details of the parameter
determination are presented. According to these findings, two black holes, nearly of
equal size of approximately 30 solar masses, merged and released an energy equal to
three solar masses. The event happened roughly at 
a distance of 400Mpc, corresponding to about 1.3 Billion light years.

One of the puzzling results is that the two black holes should have been only 
350 km apart. If these two object were formed in a binary system,
the small distance is very difficult to explain, though there are proposals
how to reach this state assuming a rapidly rotating primordial star \cite{starrot}.

Since 2009 we have proposes a modification to the Theory of General Relativity (GR),
called pseudo-complex General Relativity (pc-GR) 
\cite{hess2009,schoenenbach2012,MNRAS2013}.
Without going into details, the
main differences are: i) That a mass not only curves the space but also 
changes the vacuum properties, generating an accumulation of dark energy,
increasing toward smaller distances $R$. ii) Because of the lack of a complete quantized
theory of gravitation, a phenomenological model for the dark energy is proposed,
assuming that it increases proportional to $B/R^5$, with $R$ being the radial distance.
Finally, we require that iii) there is no event horizon, which provides a lower value
to the parameter $B$. 

Point i) is justified because semi-classical quantum mechanical calculations
reveal that vacuum fluctuations have to be increasingly present toward smaller distances
(see, for example, 
\cite{visser-boul}). However, these fluctuations explode at the Schwarzschild radius,
rendering the semi-classical approximation meaningless.
The reason is the fixed back-ground metric and a back-reaction of the dark
energy on the metric is not considered.
Point ii) is justified by i) and
that a phenomenological approach in physics is often the best way to get answers when
an approach of first principles is not available or the system is too complex. 
Proposing a definite dependence of the dark energy as a function of the radial distance,
permits to include its back-reaction on the  metric. The $R$-dependence proposed is the 
simplest one, which adds to the metric a term proportional to $1/R^3$, not entering
yet in solar system experiments \cite{will}. Point iii) has a philosophical reason which can of 
course be questioned: The event horizon is the consequence of the standard GR, however,
there is no reason to assume that it has to be so, because other properties, not considered
in GR, might come in,
introducing modifications to GR in very strong gravitational fields.
Furthermore, we find the idea not very attractive 
that, for an observer nearby, 
a region of space should be excluded from observation, thus we do not belief that the existence
of an event horizon is mandatory even if it is predicted by GR. One also has to
consider that up to now no conclusive direct observation of the event horizon
is available \cite{nohorizon}.

The parameter $B$ in the phenomenological theory is chosen such that the metric
component $g_{00}$ is always larger than zero. Thus, in this model, the large
mass concentration is rather a gray star, though in many aspects it looks like a 
black hole. This point is very delicate and, for the moment, the reader can accept it as 
a working hypothesis.

In \cite{MNRAS2013} some observational consequences were discussed, as 
of detected discrepancies to GR in galactic black holes. In \cite{MNRAS2014}
consequences to the observation of accretions disks were presented, marking relevant
differences to GR, which may be 
observationally accessible in near future \cite{horizon}. One
result of \cite{MNRAS2013}, relevant for this contribution, is that 
the orbital frequency of a point particle in a circular orbit does not increase as in GR
but near the Schwarzschild radius it deviates from the GR result, shows a maximum and
finally reaches zero at two-thirds of the Schwarzschild radius, which is according to
the model the approximate position of the surface of the star. Also, in \cite{MNRAS2014}
it is shown that the {\it Innermost Stable Circular Orbit} (ISCO), as
a function of the rotational parameter $a$, first follows the result in GR but at smaller
radial distances, until approximately for $a=0.42$ from which on there is no ISCO and stable
circular orbits exist until the surface of the star. 
In case an accretion disk is present, this produces for large $a$ an
additional bright ring in the innermost region. 
Nevertheless, the gravitational redshift increases such that approaching the surface it 
gets too large to admit a simple observation of the light emitted from the in-plunging matter.

In order to do a complete calculation of a merger, one has to recur to numerical
simulations, which we are for the moment unable to do. However, we follow the
procedure as in \cite{Ligo1,Ligo2}, where the stars are approximated by point masses 
in a circular orbit relative to each other. More details are explained in the
book by M. Maggiore \cite{maggiore}. 

The paper is organized as follows: In section 2 we resume the relevant equations
as used in GR and the differences to pc-GR. Then, the chirping mass is determined
and it is shown that within pc-GR it has to be larger than the one deduced
in \cite{Ligo1,Ligo2}. After this, the luminosity distance is obtained, which also
has to be larger. The final conclusion will be that within pc-GR the event
GW150914 corresponds to the merger of two massive black holes from
the center of two primordial galaxies which also merged before. If this is true,
it would not require a binary stellar system with two black holes at a very short
distance. The model will be kept simple and the main
objective of this contribution is to show that there may exist other 
interpretations to the source of the gravitational waves observed.

\section{The model and its consequences}

We will use the approximation of two point masses, as done in \cite{maggiore}, 
valid only as far as the two masses are not near the merging point. In pc-GR the
coupling constant depends on their separation $R$, treated as an adiabatic parameter.
Because the dependence on $R$ of the coupling constant gets pronounced at small 
distances and each star has an appreciable extension, this dependence has to be 
interpreted in average and a exact value for $R$ cannot be given.  
This will 
be the reason that near the merging point the model should work less well
and that we will present results in dependence of the parameter $R$.
Nevertheless,
the approximations will be sufficient to describe the general behavior, tendencies and to
illustrate the point we want to make.
The observer is set at a distance $r>>R$. The total mass is 
${\widetilde M}={\widetilde m}_1+{\widetilde m}_2$, with ${\widetilde m}_k$ ($k=1,2$) the 
masses of the individual participants. 
We use a tilde to denote the masses involved in pc-GR in order to distinguish to the deduced
masses within GR. 
The reduced mass ${\widetilde \mu}$ is  given by 
$\frac{{\widetilde m}_1{\widetilde m}_2}{{\widetilde M}}$.
The problem is then equivalent to the motion of a point like mass ${\widetilde \mu}$ 
around a center
with mass ${\widetilde M}$.
Because the sum of the two masses ${\widetilde m}_1$ and 
${\widetilde m}_2$ provoke the accumulation of the dark energy 
near by, the mass ${\widetilde M}$ 
is relevant, or equivalently the Schwarzschild radius 
$R_S=\frac{2G{\widetilde M}}{c^2}$.
This approximation does not include multipole contributions.

In what follows we will resume the main steps which lead to the relation of the 
chirping mass to the measured frequency and its change per unit time.
During the course of the presentation, the differences between GR and pc-GR will be emphasized.

In pc-GR a mass not only curves space but also changes the properties of the vacuum, 
accumulating dark energy with decreasing radial distance to the mass.
This part is treated in a phenomenological manner due to the lack of a full quantized GR.
For example, the $g_{00}$ component changes to \cite{hess2009}

\beqa
g_{00} & = & 1-\frac{2G{\widetilde M}}{c^2 R} + \frac{B}{2R^3}
~=~ 1-\frac{2G(R){\widetilde M}}{c^2 R} 
\nonumber \\
G(R) & = & G\left[1-\frac{b}{4R^2}\left( \frac{G{\widetilde M}}{c^2}\right)^2\right]
~=~ G\left[1-b\left( \frac{R_S}{4R}\right)^2\right]
~~~,
\label{eq1}
\eeqa
which is equivalent, as shown in the last equation, to an effective dependence of
the gravitational coupling constant on the radial distance. When the limiting
value of $B$ = $b\left( \frac{G{\widetilde M}}{c^2}\right)$, with $b=\frac{64}{27}$ is used,
the $g_{00}$ is zero at two thirds of the Schwarzschild radius (in 
order to comply with our assumption that there is no event horizon, $b$ has to be 
larger but for simplicity we use this value).

Thus, we can introduce an effective gravitational constant $G(R)  =  Gg(R)$,
where $g(R)$ acquires the form

\beqa
g(R) & = & \left[1-b\left( \frac{R_S}{4R}\right)^2\right]
~~~.
\label{eq3}
\eeqa

This has an effect on the dependence of the orbital frequency as a function
on the radial distance of a particle in a circular orbit \cite{MNRAS2014}: 

\beqa
\omega_s^2 & = & \frac{G{\widetilde M}}{R^3}F_{\omega}(R)
\nonumber \\
F_{\omega} & = & 1-\frac{3b}{4}\left(\frac{R_S}{2R}\right)^2
~~~.
\label{eq4}
\eeqa
In standard GR the factor $F_{\omega}$ is equal to one. This equation is used in turn
to depict the dependence of $R$ on the orbital frequency $\omega_s$. Here, we will do
it similar and use $F_{\omega}$ as a scaling factor.

Compared to earlier
publications we have changed to an explicit notation on masses in kg, the light
velocity and the coupling constant. Also the radial distance is now denoted as $R$,
because we distinguish between the relative distance $R$ of two masses and the
position of an observer of a gravitational wave at $r$. 
Note that in the orbital frequency
an additional factor $F_{\omega}(R)$ appears, which produces a maximum
at the distance $R=\sqrt{\frac{5B}{2R_S}}$. Toward smaller distances $R$, the orbital
frequency diminishes until, for $b=\frac{64}{27}$, it is zero at
$R=\frac{2}{3}R_S$,  which is within the model the approximate position when the two 
stars touch each other. (The sum of the radii of two equal mass gray stars, with mass
${\widetilde m}_1={\widetilde m}_2=\frac{{\widetilde M}}{2}$, is
at approximately twice the value of $\frac{2}{3}\left(\frac{R_S}{2}\right)$.)

When we consider
two point-like masses ${\widetilde m}_1$ and 
${\widetilde m}_2$, orbiting around each other at a relative 
distance $R$, the mass moments \cite{maggiore} are modified within pc-GR to

\beqa
{\widetilde M}_{ij} & = & 
{\widetilde \mu} x_i x_j
~=~ 
{\widetilde \mu} x_i x_j
~~~,
\label{eq5}
\eeqa
where the center of mass motion has been set to the origin.

The two additional factors $g(R)$ (which enters via the factor $G(R)$ in the amplitude)  
and $F_{\omega}(R)$ will be responsible 
for modifying the key equations in obtaining the chirping mass and 
the luminosity distance.

From (\ref{eq4}) we obtain

\beqa
R^2 & = & \left( \frac{G{\widetilde M}}{\omega_s^2}\right)^{\frac{2}{3}}
\left[F_{\omega}(R)\right]^{\frac{2}{3}}
\label{eq10}
\eeqa
and the amplitudes of the gravitational waves change to

\beqa
h_+(t,\theta , \phi ) & = & \frac{4G{\widetilde \mu}\omega_s^2R^2}{rc^4}
g(R) \frac{1+{\rm cos}^2(\theta )}{2} {\rm cos}(2\omega_s t_{{\rm ret}}+2\phi )
\nonumber \\
h_{\times} (t,\theta , \phi ) & = & \frac{4G{\widetilde \mu}\omega_s^sR^2}{rc^4}
g(R) {\rm cos}(\theta ) {\rm sin}(2\omega_2 t_{{\rm ret}}+2\phi )
~~~,
\label{eq11}
\eeqa
with $t_{{\rm ret}}$ is the retarted time and $g(R)$ enters via $G(R)$. 
This expression differs from
the GR result \cite{maggiore} by the factor $g(R)$. When the redshift
is taken into account and the expansion of the universe, the $r$ is changed
to the {\it luminosity distance} ${\widetilde d}_L$ and to redshifted masses.
\cite{maggiore}.

To complete the first part, we resume the result for the orbital frequency:
Using (\ref{eq10}), we get 

\beqa
E_{{\rm orbit}} & = & -\frac{G(R){\widetilde m}_1{\widetilde m}_2}{2R} 
~=~ -\frac{G{\widetilde m}_1{\widetilde m}_2}{2}
\left( \frac{\omega_s^2}{G{\widetilde M}}\right)^{\frac{1}{3}}
F_{\omega}^{-\frac{1}{3}}(R)g(R)
\nonumber \\
E_{\rm orbit} & = & 
\left( \frac{G^2{\widetilde m}_1^3{\widetilde m}_2^3\omega_s^2}{8{\widetilde M}}\right)^{\frac{1}{3}}
F_{\omega}^{-\frac{1}{3}} g(R)
~~~,
\label{eq12}
\eeqa
or

\beqa
E_{{\rm orbit}} & = & -\left( \frac{G^2\omega_{{\rm gw}}^2 
{\widetilde {\cal M}}_c^5}{32}
\right) ^{\frac{1}{3}}F_{\omega}^{-\frac{1}{3}}(R) g(R)
~~~.
\label{eq13}
\eeqa 
${\widetilde {\cal M}}_c$ = ${\widetilde \mu}^{\frac{3}{5}}{\widetilde M}^{\frac{2}{5}}$
is the chirping mass \cite{maggiore} and the frequency of
the gravitational wave $\omega_{{\rm gw}}=2\omega_s$.

Using an average ($\langle ... \rangle$) over the time, assuming that the rate of change in
time of the bacground metric is small compared to the frequency of the gravitational wave,
the energy loss per unit time and solid angle is given by

\beqa
\frac{dE}{dtd\Omega} & = & \frac{c^3 r^2}{16\pi G g(R)} \langle h_+^2 + h_{\times}^2\rangle
\nonumber \\
& = &
\frac{2c^5}{\pi G} \left( \frac{G{\widetilde {\cal M}}_c\omega_{{\rm gw}}}{2c^3}\right)^{\frac{10}{3}}
g(R)F_\omega^{\frac{4}{3}}(R)
\label{eq14}
\eeqa
and using $\frac{dE}{dt}=-\frac{dE_{{\rm orbit}}}{dt}$,
we obtain the time derivative of the circular frequency of the gravitational wave
\cite{maggiore}

\beqa
\frac{2}{3}\omega_{{\rm gw}}^{-\frac{1}{3}}\frac{d\omega_{{\rm gw}}}{dt} & = &
\frac{32}{5}\frac{c^5}{G}\left(\frac{G{\widetilde {\cal M}}_c
\omega_{{\rm gw}}}{2c^3}\right)^{\frac{10}{3}}
\left( \frac{32}{G^2{\widetilde {\cal M}}_c^5}\right)^{\frac{1}{3}}
F_{\omega}^{\frac{5}{3}}(R)
\nonumber \\
& = &
\frac{12}{5}2^{\frac{1}{3}}\left( \frac{G{\widetilde {\cal M}}_c}{c^3}\right)^{\frac{5}{3}}
\omega_{{\rm gw}}^{\frac{11}{3}}
F_{\omega}^{\frac{5}{3}}
~~~.
\label{eq15}
\eeqa

Using for the frequency $f_{{\rm gw}} = \frac{\omega_{{\rm gw}}}{2\pi}$, we obtain

\beqa
\frac{df_{{\rm gw}}}{dt} & = &
\frac{96}{5} \pi^{\frac{5}{8}}
\left( \frac{G{\widetilde {\cal M}}_c}{c^3}\right)^{\frac{5}{3}}f_{{\rm gw}}^{\frac{11}{3}}
F_{\omega}^{\frac{5}{3}}(R)
\label{eq16}
\eeqa
and solving for the chirping mass, as it appears in {\cite{Ligo1,Ligo2},
shifting the additional factors to the side of the chirping mass, we obtain

\beqa
{\cal M}_c ~=~
{\widetilde {\cal M}}_c F_{\omega}(R)
& = &
\frac{c^2}{G}\left[ \frac{5}{96\pi}{\frac{8}{3}} \frac{df_{{\rm gw}}}{dt} 
f_{{\rm gw}}^{-\frac{11}{2}}
\right]^{\frac{3}{5}}
~~~.
\label{eq17}
\eeqa
Note, that the factor $F_{\omega}$
on the left decreases with lower distances $R$ and that the value on the
right is obtained from the observed frequency and its change in time 
of the gravitational wave. If (\ref{eq17}) is
of the order of ${\cal M}_c=30$ \cite{Ligo1}
(we are not taking the exact deduced values because we are only interested in
the general behavior), the chirping mass ${\widetilde {\cal M}}_c$ on the left has to be larger.
When the factor approaches zero, the chirping mass tends to infinity. Of course, our
model of two point masses, orbiting each other, and the approximation of an effective
gravitational constant, has to be taken with care and very probably is inaccurate near
the merging point of the two masses. Nevertheless, the result shows that within pc-GR the
chirping mass is probably larger and does not correspond to two stars with 30 solar masses 
each.

Using for $b$ the limiting values, the factors $g(R)$ and $F_{\omega}(R)$ have the form

\beqa
g(R) & = & 1-\frac{1}{3}\left( \frac{2R_S}{3R}\right)^2
\nonumber \\
F_{\omega}(R) & = & 1-\left(\frac{2R_S}{3R}\right)^2
~~~.
\label{eq18}
\eeqa

In a second step, we determine which value the luminosity distance has to acquire,
without contradicting the observation, i.e., we will {\it not derive} its value
but rather present an estimation:

Assuming a flat universe and the evolution of the dark energy as a function
in time is equal to the present best known models (in \cite{maglahui} other scenarios
are discussed), the luminosity distance can be determined as a function of the redshift in a
standard manner (using $\Omega_0=0.3036$ and ${\rm H}_0=68.14$ \cite{cal}).
Here, we are interested to estimate the luminosity distance required in order 
to get the {\it same amplitude for the gravitational wave}.
Note, that the amplitude of the gravitational wave is given by (see Eq. (\ref{eq11}))

\beqa
A & = & 
\frac{4G{\widetilde \mu}\omega_s^2R^2}{r c^4}g(R)
~~~.
\label{eq19}
\eeqa
Substituting $r$ by the luminosity distance ${\widetilde d}_L$, this translates to

\beqa
A & = & \frac{4G^{\frac{5}{3}}}{c^4}
\frac{{\widetilde \mu}{\widetilde M}^{\frac{2}{3}}\omega_s^{\frac{2}{3}}}{{\widetilde d}_L}
F_{\omega}^{\frac{2}{3}}(R)g(R)
\nonumber \\
& = &
\frac{4G^{\frac{5}{3}}}{c^4} 
\frac{\widetilde {\cal M}_c^{\frac{5}{3}}\omega_s^{\frac{2}{3}}}
{{\widetilde d}_L}F_{\omega}^{\frac{2}{3}}(R) g(R)
~~~.
\label{eq20}
\eeqa

Demanding that in pc-GR the same amplitude is reproduced as in GR,
leads to the relation 

\beqa
\frac{{\widetilde {\cal M}}_c^{\frac{5}{3}}}{{\widetilde d}_L} F_{\omega}^{\frac{2}{3}}(R) g(R)
& = &
\frac{{\cal M}_c^{\frac{5}{3}}}{d_L}
~~~,
\label{eq21}
\eeqa
where $d_L$ is the luminosity distance as deduced in \cite{Ligo1,Ligo2}. 
Using (\ref{eq17}) and resolving for ${\widetilde d}_L$, leads finally to

\beqa
{\widetilde d}_L & = & d_L \frac{g(R)}{F_{\omega}(R)}
~~~.
\label{eq22}
\eeqa
Note, that the function $g/F_{\omega}$ becomes very large for 
smaller relative distances
$R$ of the two massive objects.

\begin{figure}[ht]
\begin{center}
\includegraphics[width=12cm,height=20cm]{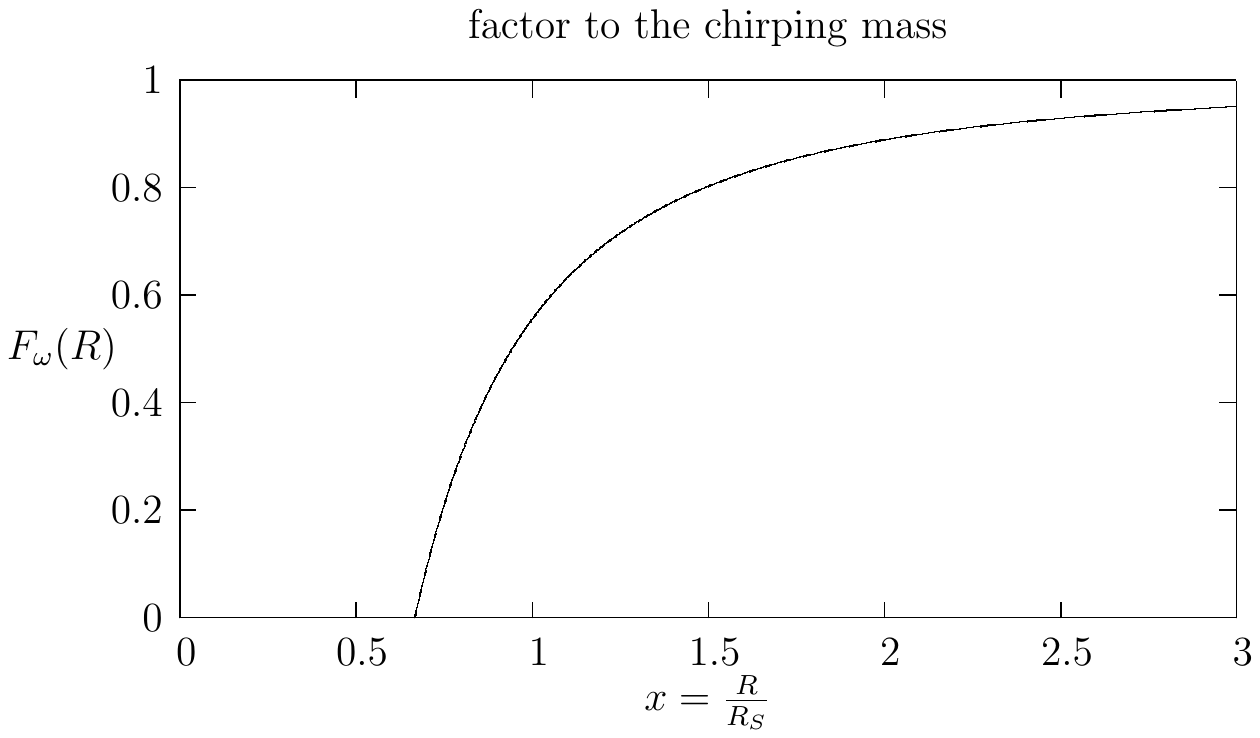}
\vskip -12cm
\caption{The factor to the chirping mass in Eq. (\ref{eq17}).}
\end{center}
\label{fig1}
\end{figure}

The inverse of $F_{\omega}(R)$, plotted in Fig. \ref{fig1}, gives the factor by which the 
reported value ${\cal M}_c$ has to be multiplied
in order to obtain ${\widetilde {\cal M}}_c$.
For a given value of $x=\frac{R}{R_S}$ a corresponding new chirping mass and a luminosity
distance ${\widetilde d}_L$ in Eq. (\ref{eq22}) is obtained.

For example, for 
$x \approx 0.6682$ the factor for the chirping mass is approximately 218, i.e. a new 
chirping mass of about ${\widetilde M}_c = 6540$. The luminosity distance
is then ${\widetilde d}_L$ = $400$~Mpc * $146$ = $58400$, which corresponds in 
the standard model of a flat universe to a redshift of a little less than $z=6$,
the limit where the first galaxies may have formed (in \cite{galaxies} galaxies
with a redshift between 4 to 4.5 have been observed).
Remember that the two massive objects touch (merge) at $x=2/3=0.66$.
For larger $R$, the final masses decrease and also the value of $z$. For smaller 
relative distances the masses increase to very large values, however the redshift factor 
is then too large, such that galaxies did not exist.

One has also to keep in mind that the model presented is very simple and larger masses
at moderate $z$-values cannot be excluded.

\begin{table}
\begin{center}
\begin{tabular}{|c|c|c|c|}
\hline
$x=\frac{R}{R_S}$ & $F(R)$ & {\rm factor~for~} ${\widetilde d}_L$ & $z$ ({\rm approx.})\\
\hline
0.6682  & 0.00458 & 145.8 & $\approx$ 6\\
0.6690  & 0.00667 & 96.1 & $\approx$ 4 \\
0.7000  & 0.09297 & 7.5 & $\approx$ 0.5 \\
\hline
\end{tabular}
\end{center}
\caption{\label{tab1} Sample values for three distance values $x=\frac{R}{R_S}$, where
$R_S$ is the Schwarzschild radius for the combined mass ${\widetilde M}$. The
first column lists the values of $x$. The second column tabulates the factor 
$F_{\omega}(R)$ by which
the chirping mass ${\widetilde {\cal M}}_c$ in pc-GR is multiplied in order to obtain the
observational deduced value ${\cal M}_c$. The third column lists the factor by which
the experimental deduced luminosity distance $d_L=400$~Mpc has to be multiplied in order
to obtain the luminosity distance ${\widetilde d}_L$ within pc-GR. The last column gives
the approximate redshift factor $z$.}
\end{table}

Some sample values for different distances $x=\frac{R}{R_S}$ are listed in Table 
\ref{tab1}.

\section{Conclusions}

We have applied a modified theory of GR, called pc-GR, which adds near great masses
a dependence on a dark energy, acting repulsively. We resumed the 
differences of the equations in pc-GR compared to GR,
used to obtain the amplitude of the gravitational wave and the chirping mass. The main result is
that the chirping mass is larger than the reported one in \cite{Ligo1,Ligo2}
and that the luminosity distance corresponds to the remote past of the universe. 
The system, producing the gravitational wave, would then consist of two large
black holes, which were formerly at the center of two merging primordial galaxies.
This interpretation seems to us more plausible than assuming a rapidly rotating
super-massive star, which fissions in two very large chunks, undergoing at the same time
a collapse as a supernova. Also the formation of two black holes from super-massive
stars in a binary system at such short distances is questionable.
Maybe, here we have a case where the first observation of
gravitational waves hints to a needed modification of GR in very strong fields.

The model applied is very simple and results in a realistic treatment
may change such that higher masses with a smaller luminosity distance may be
possible. The present contribution has the objective to show that other possibilities
exist for the interpretation of the initial object of GW150914.

Just now, another event was reported \cite{gw2} with a chirping mass of approximately 9
solar masses, at 440~Mpc. All interpretations given here are also applicable to
this new event, i.e. that it should be a merger of two massive black holes,
following a merger of two primordial galaxies. A further possible event is reported in
\cite{gw3}, together with the two other events. Again our conlusions is the same that
one observed the merger of large black holes.

\section{Acknowledgment}
{\textsc{Peter O. Hess}} acknowledges the financial support from DGAPA-PAPIIT (IN100315).
He also acknowledges the hospitality at the {\it Frankfurt Institute for Adevances Studies},
where part of the work has been realized. The author is very grateful for comments given by
T. Boller (Max Planck Institute for Extraterrestrial Physics, Garching, Germany).
\label{sec:Ack}

\end{document}